\documentclass[tightenlines,prd,aps,nofootinbib,groupedaddress,english,eqsecnum,superscriptaddress,11pt]{revtex4-1}

\usepackage{url}
\usepackage{graphicx}
\usepackage{amsmath,amssymb,amstext,amssymb,amsfonts,amsthm}
\usepackage{hyperref}
\usepackage{appendix}
\usepackage[margin=1.0in,papersize={8.5in,11in}]{geometry}
\usepackage[utf8]{inputenc}

\usepackage{color}
\usepackage[normalem]{ulem}

\graphicspath{{plots/}}

\begin{document}

\title{On the question of measuring spatial curvature in an inhomogeneous universe}

\author{Chi Tian}
\email{chit@wustl.edu} 
\affiliation{CERCA/ISO, Department of Physics, Case Western Reserve University, 10900 Euclid Avenue, Cleveland, OH 44106, USA}
\affiliation{Department of Physics and McDonnell Center for the Space Sciences,
Washington University, St. Louis, MO 63130, USA}

\author{Stefano Anselmi}
\affiliation{Department of Physics, Israel Institute of Technology, Haifa 320003, Israel}
\affiliation{INFN, Sezione di Padova, via Marzolo 8, I-35131, Padova, Italy}
\affiliation{Observatoire de Paris, PSL Research University, Universite de Paris, 92190 Meudon, France}

\author{Matthew F. Carney}
\affiliation{Department of Physics and McDonnell Center for the Space Sciences,
Washington University, St. Louis, MO 63130, USA}

\author{John T. Giblin, Jr}
\affiliation{Department of Physics, Kenyon College, 201 N College Rd, Gambier, OH 43022, USA}
\affiliation{CERCA/ISO, Department of Physics, Case Western Reserve University, 10900
Euclid Avenue, Cleveland, OH 44106, USA}

\author{James B. Mertens}
\affiliation{Department of Physics and McDonnell Center for the Space Sciences,
Washington University, St. Louis, MO 63130, USA}
\affiliation{Department of Physics and Astronomy, York University, Toronto, Ontario, M3J 1P3, Canada}
\affiliation{Perimeter Institute for Theoretical Physics, Waterloo, Ontario N2L 2Y5, Canada}

\author{Glenn Starkman}
\affiliation{CERCA/ISO, Department of Physics, Case Western Reserve University, 10900 Euclid Avenue, Cleveland, OH 44106, USA}

\begin{abstract}
    The curvature of a spacetime, either in a topological sense, or averaged over super-horizon-sized patches, is often equated with the global curvature term that appears in Friedmann's equation.
    In general, however, the Universe is inhomogeneous, and gravity is a nonlinear theory, thus any curvature perturbations violate the assumptions of the FLRW model;  it is not necessarily true that local curvature, averaged over patches of constant-time surfaces, will reproduce the observational effects of global symmetry.
    Further, the curvature of a constant-time hypersurface is not an observable quantity, and can only be inferred indirectly.
    Here, we examine the behavior of curvature modes on hypersurfaces of an inhomogeneous spacetime non-perturbatively in a numerical relativistic setting, and how this curvature corresponds with that inferred by observers. 
    We also note the point at which observations become sensitive to the impact of curvature sourced by inhomogeneities on inferred average properties, finding general agreement with past literature.
    
\end{abstract}

\maketitle

\section{Introduction}

Much of our understanding of cosmology relies upon the {\sl cosmological principle}: our Universe is homogeneous and isotropic on large scales.
This assumption gives rise to the (perturbed) Friedmann-Lema\"itre-Robertson-Walker (FLRW) model, a dynamical model of the Universe in general relativity (GR), in which background quantities are homogeneous and independent of the perturbations, and gravitational perturbations are, typically, only treated at first order.
Cosmological surveys generally quantify
the nature of the contents and other characteristics of our Universe by 
inferring the expansion history of the background cosmology.
While 
such efforts have placed tight constraints on
the average abundances of various stress-energy components \cite{Riess:2019cxk,Aghanim:2018eyx},
discrepancies among measurements
have arisen, including the local rate of expansion---the Hubble tension (see e.g. \cite{2007.10716} for a recent review), 
as well as the amount of matter and amplitude of inhomogeneities (see e.g. \cite{1901.05289}).
One common way to resolve these cosmological tensions is to introduce new physics---additional parameters that restore our 
confidence in the FLRW model. 
Other proposals have invoked new phenomenology within existing physics, such as the (re-)introduction of spatial curvature as an important contribution to the evolution of the Universe \cite{1712.02967,1902.07915,2002.10831,1903.11433}.

While FLRW models are built upon an assumption of homogeneity, our Universe is not perfectly so. When fitting our cosmological observations
to an FLRW model, we are seeking best-fit homogeneous parameters in an inhomogeneous universe, giving rise to questions including the following:
What is the best fit FLRW model?
How do we evaluate goodness of this fit? 
Will the inhomogeneities introduce biases in our evaluations of FLRW parameters? 
These questions are closely related to the so-called ``fitting problem'' \cite{Ellis_1987}, and are
challenging to study within the traditional cosmological framework.
By fitting 
cosmological observables generated by general relativistic simulations -- simulations not reliant upon an FLRW model -- with the FLRW model,
we can nevertheless explore those questions, and better understand how and when fitting-related issues arise.

Many studies draw parallels between the curvature parameter, $\Omega_k\equiv  k / H_0^2$---which is a phenomenological fitting parameter in the Friedmann equation---and
the Ricci curvature parameter $\Omega_R \equiv \left<R\right> / (6 H_0^2) $,
which is proportional to the spatial average of the Ricci tensor 
on spatial hypersurfaces \cite{1202.5037,1707.01800,1708.09143,1205.2422,2002.10831,Rasanen:2005zy,0801.2692,Koksbang:2019cen,Koksbang:2019glb}. 
In a homogeneous universe these are identical; the usual thought is that in an inhomogeneous universe, they should also be approximately equal.
However, the critical point here is that the average Ricci curvature on a hypersurface {\em does not actually correspond to an observable, nor to the inferred cosmological parameter $\Omega_k$}.
It is then interesting to ask to what extent the observed curvature can be used to describe properties of spatial hypersurfaces. 
Measuring and interpreting $\Omega_k$ properly will be necessary to understand and characterize physics that can produce curvature, for example non-trivial cosmic topology 
\cite{LachiezeRey:1995kj,Levin:2001fg,Gausmann:2001aa,Gausmann:2001aa,Riazuelo:2002ct,Lehoucq:1998yu,Cornish:1997hz,Vaudrevange:2012da} and
some  inflationary models \cite{1202.5037,astro-ph/0503405,0804.1771,0901.3354}.

To {\sl measure} $\Omega_k$, one could look to
Hubble diagrams obtained from observations of standard candles such as
type Ia supernovae, to 
a statistical analysis of the Cosmic Microwave Background (CMB) or of measurements of the Baryon Acoustic Oscillations (BAO) \cite{Aghanim:2018eyx,1907.01033,0712.3457,0811.0981,1711.03437,1710.00845}, or to a variety of other observations of tracers at cosmological distances.  All of these observations rely on measuring photons, which follow null geodesics of the perturbed FLRW metric. The past null geodesics of an observer define a light cone, a null hypersurface embedded in a 4-dimensional manifold.
However, analytic work describing our Universe
is commonly written in a 3+1 formalism, which foliates
a 4-dimensional manifold with spacelike, constant-time, rather than null, hypersurfaces.
In an exactly FLRW universe one can use null-hypersurfaces to measure cosmological parameters,  however, one cannot draw on the same assumptions in an inhomogenous one.

In this study, we will employ the most direct methodology for understanding how an inhomogeneous 
and presssureless matter fluid  can affect measurements of the expansion history of the Universe---we will numerically solve the Einstein equations 
for cosmological fluid perturbations alongside the geodesic equations for photons that will probe the Universe.
While similar efforts have been made to examine the accuracy of the FLRW model by including general relativistic effects 
\cite{East:2017qmk, Giblin:2017ezj, Bentivegna:2015flc,Bentivegna:2016fls, Giblin:2015vwq, Macpherson:2018btl,Macpherson:2018akp,Giblin:2018ndw, Meures:2011gp, Adamek:2018rru}, cosmological observables, especially $\Omega_k$, 
have not been thoroughly explored non-perturbatively in a full numerical GR setting.
We construct Hubble diagrams from numerical-relativistic simulations of a universe with collisionless matter and a cosmological constant, to examine 
the accuracy of the FLRW model and extract any possible bias from inhomegeneities. The cosmological perturbations are modeled as single
modes of variable amplitude whose wavelengths are also varied from super-horizon size to less than $10 \, \rm{Mpc}$. 
We find that no detectable bias on FLRW parameters is introduced  by fitting FLRW models.
We also investigate and show the difference between $\Omega_{k}$ inferred from fitting the Hubble diagram, the Ricci curvature parameter $\Omega_R$ on the past lightcone, and the $\Omega_R$ on spatial hypersurfaces.

We begin in Section \ref{sec:method} by introducing the numerical 
relativity formalism and code that  we use to evolve the metric and matter fields.
A brief introduction to relativistic ray tracing and the 
method to extract the distance measure are also presented.
We present our main results in Section \ref{sec:res}, which 
discusses the possible bias introduced by the cosmic modes on 
matter fields with amplitudes given by standard cosmological 
perturbation theory. We compare different curvature measures mentioned above and confirm that 
the observed $\Omega_k$ depends neither on the curvature averaged over the spatial hypersurface nor its average over the light cone.

\section{Methodology}
\label{sec:method}

We study a universe starting at redshift $z= 5$ containing perturbed collisionless matter and a cosmological constant,
with respective homogeneous density parameters 
that eventually yield 
$\Omega_m = 0.3 $ and $\Omega_{\Lambda} = 0.7$
at the end of the simulation when $z = 0$ and mock observations are generated.
The collisionless matter is modeled by a pressureless, $w=0$, fluid coupled to gravity, whose influence to the Universe expansion history has been studied by \cite{Giblin:2015vwq}. 
The dynamics of the spacetime is solved for using the BSSNOK formulation
of numerical relativity \cite{Nakamura:1987zz,Shibata:1995we,Baumgarte:1998te},
which is a reformulation of Einstein's equations to a $3+1$/ADM form based on a conformal 
transformation of the metric. 
This reformulation permits use of an arbitrary gauge and stress-energy source,
allowing us to investigate the evolution of matter fields in a cosmological setting
for an arbitrary gauge choice, or slicing condition. 
This is especially important for computing the spacetime behavior in comoving-synchronous gauge, 
or geodesic slicing: this is the coordinate system in which fluid observers are at rest and have experienced a fixed passage of proper time.
We will preferentially work in this gauge,
as the fluid provides us with a natural set of observers, and this gauge provides us with access to their rest frames.

The BSSNOK formulation 
parametrizes a general metric as
\begin{equation}
  ds^{2}=-\alpha^{2}dt^{2}+\gamma_{ij}\left(dx^{i}+\beta^{i}dt\right)\left(dx^{j}+\beta^{j}dt\right)\,,
\end{equation}
where $\gamma_{ij}$ is the spatial metric defining the intrinsic geometry on the spatial hypersurfaces, and $\alpha$ and $\beta^i$ are the
lapse and shift respectively.
In the comoving synchronous gauge these are
\begin{align}
\label{eq:5}
\alpha = 1,\;\;\beta = 0.
\end{align}
In this gauge, because the coordinates are comoving with the pressureless dust fluid, the fluid will have zero  velocity.
This allows us to cast the fluid equations of motion into the simple conservative form,
\begin{align}
\label{eq:1}
\partial_t \tilde{D} \equiv \partial_t \left( \gamma^{1/2} \rho \right) = 0,
\end{align}
where $\rho$ is the projection of the stress tensor along some normal direction $n^a$
\begin{align}
    \rho = n^a n^b T_{ab},
\end{align}
and $\gamma$ is the determinant of the induced 3-metric $\gamma_{ij}$. The scalar curvature of the spatial hypersurface, ${}^3R$, corresponds to the 3-metric $\gamma_{ij}$, and we refer to this throughout this work without the preceding superscript.

A single mode is superimposed onto
this homogeneous background. To set-up the initial conditions for the metric degrees of freedom
on an initial spatial hypersurface, we follow the
method employed in \cite{Mertens:2015ttp}, 
 conformally transform $\gamma_{ij} = \psi^4 \tilde{\gamma}_{ij}$,
and then decompose
$\rho$ into two pieces, $\rho_K$ and $\rho_{\psi}$,
which source the trace of the extrinsic curvature 
$K$ and the conformal factor $\psi$ respectively.
Employing the condition of conformal flatness $\tilde{\gamma}_{ij} = \delta_{ij}$,
the Hamiltonian constraint is reduced to
\begin{align}
  \label{eq:2}
    K & = - \sqrt{24\pi \rho_K}\\
  \nabla^2 \psi &= -2 \pi \psi^5 \rho_{\psi},
\end{align}
and the momentum constraint is manifestly satisfied. While solving the first of these equations is straightforward,
the second equation involves a non-linear term that makes it
difficult to solve for $\psi$ a given $\rho_{\psi}$. 
Instead we consider simple forms of $\psi$ and calculate $\rho_\psi$.
We also focus on the case with planar symmetry with periodic boundary conditions to reduce 
the full problem to a 1-dimensional problem. 
Specifically, we take $\psi = 1 + \Delta \psi$
where $\Delta \psi$ is a single mode
\begin{align}
\label{eq:3}
\Delta \psi = A \sin (2\pi x/L),
\end{align}
for the simplified 1D problem. Note that although the single mode model cannot represent the real universe, it disentangles the inhomgeneities mode by mode so that will reveal the most relevant mode length that brings possible biases.

To construct the Hubble diagrams for comoving observers at different position,
we start by evolving the matter and geometry from the initial time to $z=0$. We then propagate rays backward in time to $z = 1$ through the simulation starting (i.e. ending) at observers uniformly distributed along the $x$ axis. 
For each observer, we choose to shoot
light rays uniformly distributed
in solid angle in its own inertial frame. 
Because of the planar symmetry, the spherically symmetric rays 
can be projected on the 2-dimensional $x-y$ plane, 
and 
the 3D distribution of rays can be restored by assigning 
weights to each ray on the 2D plane.
Since the distribution of rays is uniform in the observer's rest frame,
which is locally Minkowskian,
it is  necessary to transform them into the comoving 
frame in the configurations of the simulations.

The ray-tracing is accomplished by solving the geodesic equation 
for photons
\begin{align}
p^a \nabla_a p_b =0,
\end{align}
where the four-momentum $p^a$ has components
\begin{align}
    p^i &= dx^i / ds \nonumber \\
    p^0 &= \frac1\alpha \sqrt{\gamma^{ij} p_i p_j},
\end{align}
and $s$ is an affine parameter.
To incorporate the 3+1 form to the geodesic equations, following \cite{PhysRevD.49.4004}, we can rewrite it as
\begin{align}
    \frac{dp_i}{d s} &= \alpha \alpha_{,i} (p^0)^2 + \beta_{,i}^k p_k p ^0
    - \frac12 \gamma_{,i}^{lm}p_l p _m \\
    \frac{dx^j}{d s} &= \gamma^{ij} p_i - \beta^j p^0, \\
    \frac{ds}{dt} &= (p^0)^{-1}
\end{align}
The redshift is then defined 
as 
\begin{align}
    z = \frac{u_S^{\mu} p_{\mu}}{u_O^{\mu} p_{\mu}}-1,
\end{align}
where $u_S^{\mu}$ and $u_O^{\mu}$ are four velocities of the source and 
the observer respectively and are $u_{S,O}^\mu = (1,0,0,0)$ for comoving observers.

To calculate the luminosity distance, for every light ray received by each observer, we evolve two additional rays that slightly deviate from the original ray.
Their deviations are orthogonal to one another and to the momentum of the original ray, and these orthogonal directions are chosen by employing the Gram-Schmidt 
process with the original direction $p_{\mu}$ and  two other arbitrary non-parallel vectors $s_1^{\mu}$ and $s_2^{\mu}$.
After applying the Gram-Schmidt algorithm,  $s_1^{\mu}$ and $s_2^{\mu}$ are cast into orthogonal basis $\hat{s}_1$ and 
$\hat{s}_2$.
The two auxiliary  rays are then generated that are pointing to $p_{\mu}+ \epsilon \hat{s}_1$ and $p_{\mu} + \epsilon \hat{s}_2$ respectively, where 
the infinitesimal parameter $\epsilon$ controls the width between two adjacent rays. 
 $\epsilon$ is chosen to be sufficiently small so that the distance-redshift relation is independent of $\epsilon$.
Distance measures can be 
identified through tracing the shape of infinitesimal triangles formed by these three-ray combinations on a screen plane;
we present details in Appendix \ref{appd:2}.

\section{Results}
\label{sec:res}
To ensure a statistically similar distribution of matter as in the real Universe, 
we choose the value of the initial
amplitude of the mode, $A$, to correspond to 
the rms density perturbation
at $z=5$ smoothed on a length scale $L$,
\begin{equation}
\label{eq:sigma_rho}
\frac{ \sigma_{\rho, L}^2 }{ \bar{\rho}^2 } = \frac{1}{2\pi^2} \int k^2 e^{-(k L)^2} P_{\delta\delta}(k) dk\,.
\end{equation}
The comoving-synchronous-gauge matter-density power spectrum $P_{\delta\delta}(k)$ \cite{Lewis:2002ah}
is calculated using CAMB \cite{Lewis_2000}, while $\bar{\rho}$ can 
be identified as $\rho_K$ in Eq. \eqref{eq:2}.
We use CAMB settings including the Halofit nonlinear matter power spectrum,
$H_0 = 67.5$ km/s/Mpc, $\Omega_b h^2 = 0.022$, $\Omega_c h^2=0.122$, $n_s = 0.965$,
and otherwise default settings to compute $\sigma_{\rho, L}$.

The trace of the extrinsic curvature $K$ describes the local expansion rate, which connects to the effective Hubble parameter $H$ in the FLRW limit through $H = K / 3$.  Its initial value can be determined from $\rho_K$ using Eq. \eqref{eq:2}, and thus, it also determines the initial Hubble parameter $H_I$. The simulation employs geometric units, and the mass scale $M$ can be fixed by comparing the numerical value of $K$ at $z = 0$ with
the current Hubble scale, $H_0^{-1} \sim 4.4\; \mathrm{Gpc}$. 
After fixing the Hubble length, we choose our box sizes to  
vary from as large as $5$ times the Hubble scale $H_I^{-1}$ at 
$z = 5$ to $0.005 H_I^{-1}$,
and calculate the corresponding distance-redshift relation.

For each observer, we first eliminate data points with redshifts less than $z=0.03$, then we calculate the {\sl angle}-averaged luminosity distance, $\left<D_L\right>$;
we randomly sample $N=1000$ data points and apply a least-square fit to their 
distance-redshift relation with
the Levenberg–Marquardt algorithm.
We chose enough points so that the fit converges by $N=1000$ points.

The distance-redshift relation $D_L(z)$ assuming an FLRW model can be written as 
\begin{equation}
\label{eq:dl}
D_L(z) = 
\begin{cases}
\frac{1+z}{H_0}|\Omega_k|^{-1/2} \sinh{\left[|\Omega_k|^{1/2}H_0 \chi(z)\right]}, & \Omega_k > 0,\\
(1+z)\chi(z), & \Omega_k = 0, \\
\frac{1+z}{H_0}|\Omega_k|^{-1/2} \sin{\left[|\Omega_k|^{1/2}H_0 \chi(z)\right]}, & \Omega_k <0.
\end{cases}
\end{equation}
where $\chi(z) = \int 1/H(z) dz$ is the comoving distance to redshift $z$
and 
the redshift dependence of $H(z)$ depends on the specific matter components 
we include in the FLRW model.
Although the only matter fields 
we have added in our simulation are matter and dark energy, we will try to fit
FLRW models with different matter components to examine any effective 
matter components that are fitting consequences.
The FLRW models we investigate are listed in Table \ref{t1}, 
and the constraint $\Sigma_i \Omega_i = 1 $ is enforced by fixing 
one of the matter components. 

\begin{table}[htbp]
  \caption{\label{t1} Parameters of initial setups. For each model, 
  there is one energy component that is fixed by the requirement 
  $\Sigma_i \Omega_i = 1 $. }
  \begin{center}
  \item[]\begin{tabular}{c|c}
    \hline
    \hline
    Models & Parameters\\
    \hline 
    M1 & $H_0$, $\Omega_m = 1$ \\
    M2 & $H_0$, $\Omega_m$, $\Omega_{\Lambda} = 1-\Omega_m$ \\
    M3 & $H_0$, $\Omega_m$,  $\Omega_{\Lambda}$, $\Omega_k=1-\Omega_m - \Omega_{\Lambda}$ \\
    M4 & $H_0$, $\Omega_m$, $\Omega_{\Lambda}$,  $\Omega_r$, $\Omega_k = 1-\Omega_m - \Omega_{\Lambda} - \Omega_r$
  \end{tabular}       
  \end{center}
\end{table}

\subsection{The accuracy of the FLRW model}

\begin{figure}[htbp]
\centerline{\includegraphics[width=0.7\textwidth]{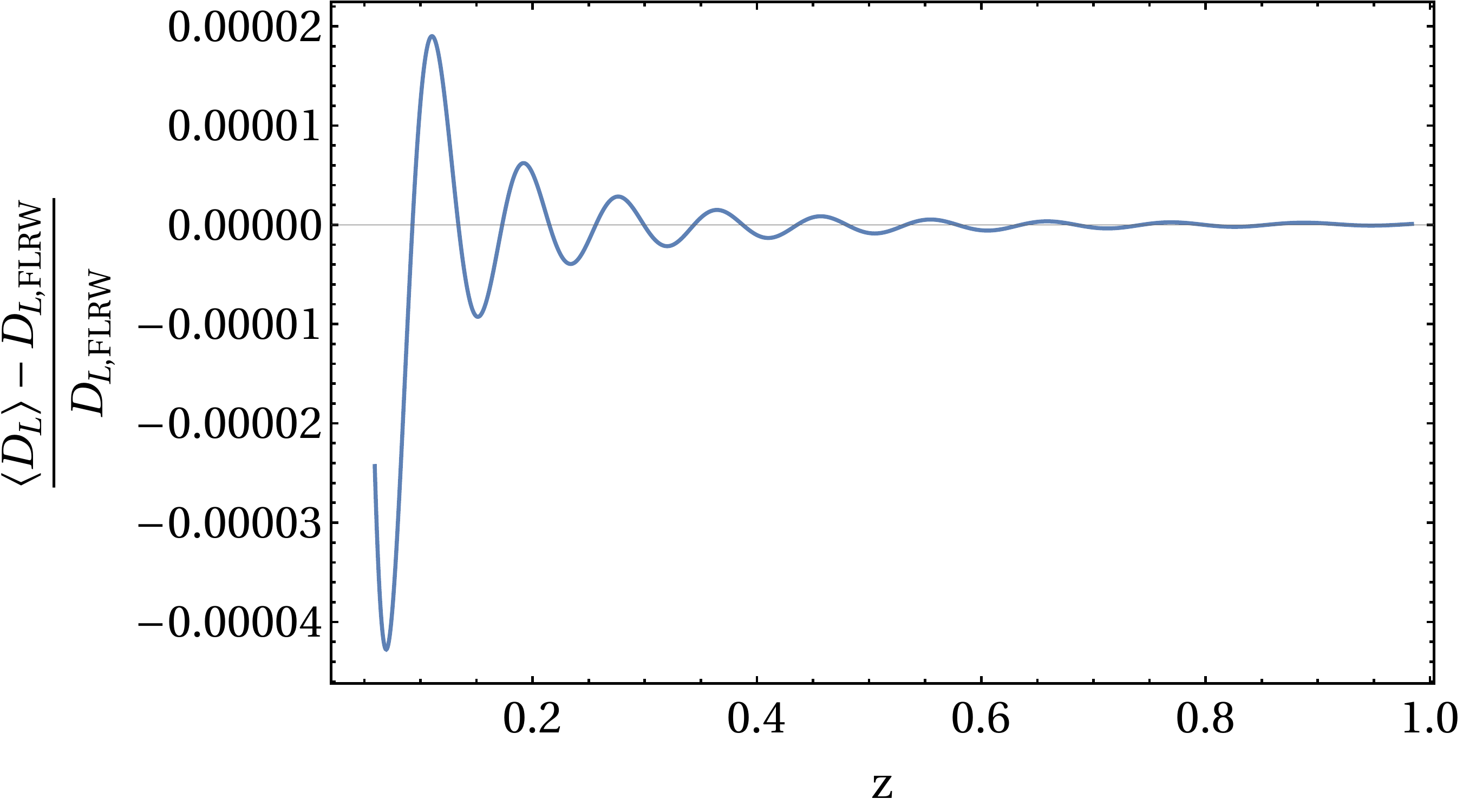}}
\caption[]{\label{fig:avg} Comparing the angle-averaged $\left<D_L(z)\right>$ 
to the best fitted FLRW distance-redshift relation $D_{L,\rm FLRW} (z)$. The initial mode has $\lambda = 0.1 H_I^{-1}$, and the observer is at the density global maximum.
}
\end{figure}

The inhomogeneity of the universe will introduce fluctuations on top of the distance-redshift relations of the FLRW model, as shown in Fig. \ref{fig:avg} 
for a comoving observer at $z = 0$ located 
at a position where the density field has a global maximum. 
Fig. \ref{fig:avg} compares the angle-averaged $\left<D_L\right>$ of all the data points to the best-fit 
FLRW model with only $\Omega_m$ and $\Omega_L$ (model M2 in Table \ref{t1}). 
Even though the deviations are small, the oscillatory feature is clear.

Because of the deviations from homogeneity,
questions arise when using the homogeneous FLRW 
model to fit this inhomogeneous universe---what is the best fit FLRW model and how good is it?
To evaluate the goodness of fit of the different FLRW models, we add constant mock errors to each data point and 
estimate the minimum error needed to make the fit trustworthy by
evaluating the resulting p-value for each model under the hypotheses that the data is consistent with the model. 
For a given level of mock error, a higher p-value indicates a better fit. 
A p-value  $p=.05$ is chosen as the threshold for an acceptable fit, which corresponds to a chi-squared-per-degree-of-freedom that is close to 1.  We pick the observer that has the worst FLRW fit (smallest p-value),
which turns out to be the observer locating at the global density peak.
The results for the minimum errors needed to validate the fittings are shown 
in Table \ref{tab:error_table}.
\begin{table}[htbp]
  \includegraphics[width=0.95\textwidth]{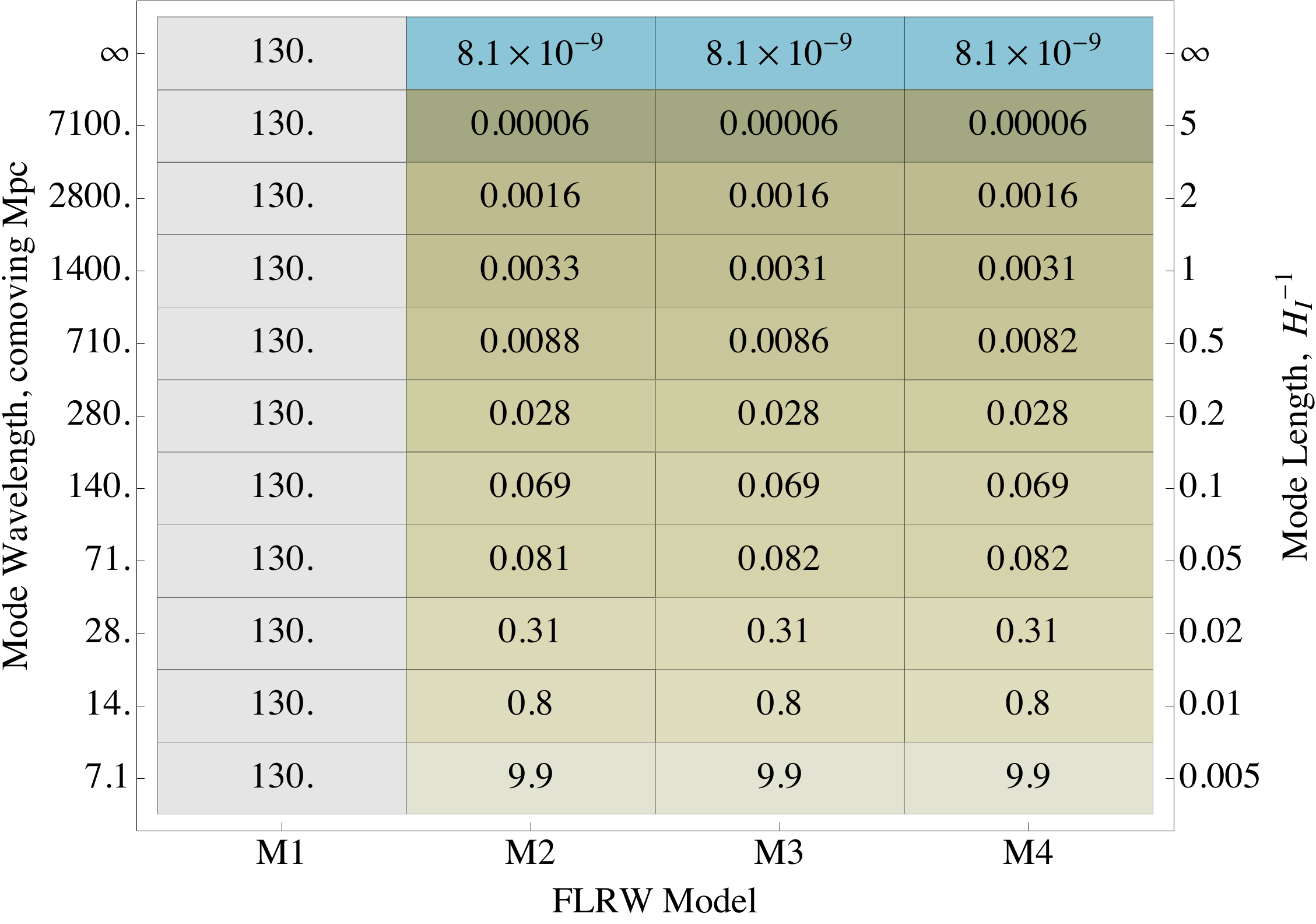}
    \caption{For each combination of the FLRW model and mode size, 
  the minimum constant mock errors needed in comoving $\rm Mpc$ to give fitting p-value $p \sim 0.05$. The mode, $\lambda=\infty$, is a homogeneous simulation, where the FLRW symmetry is exact.
  \label{tab:error_table}
  }
\end{table}

\begin{figure}[htbp]
\centerline{\includegraphics[width=0.7\textwidth]{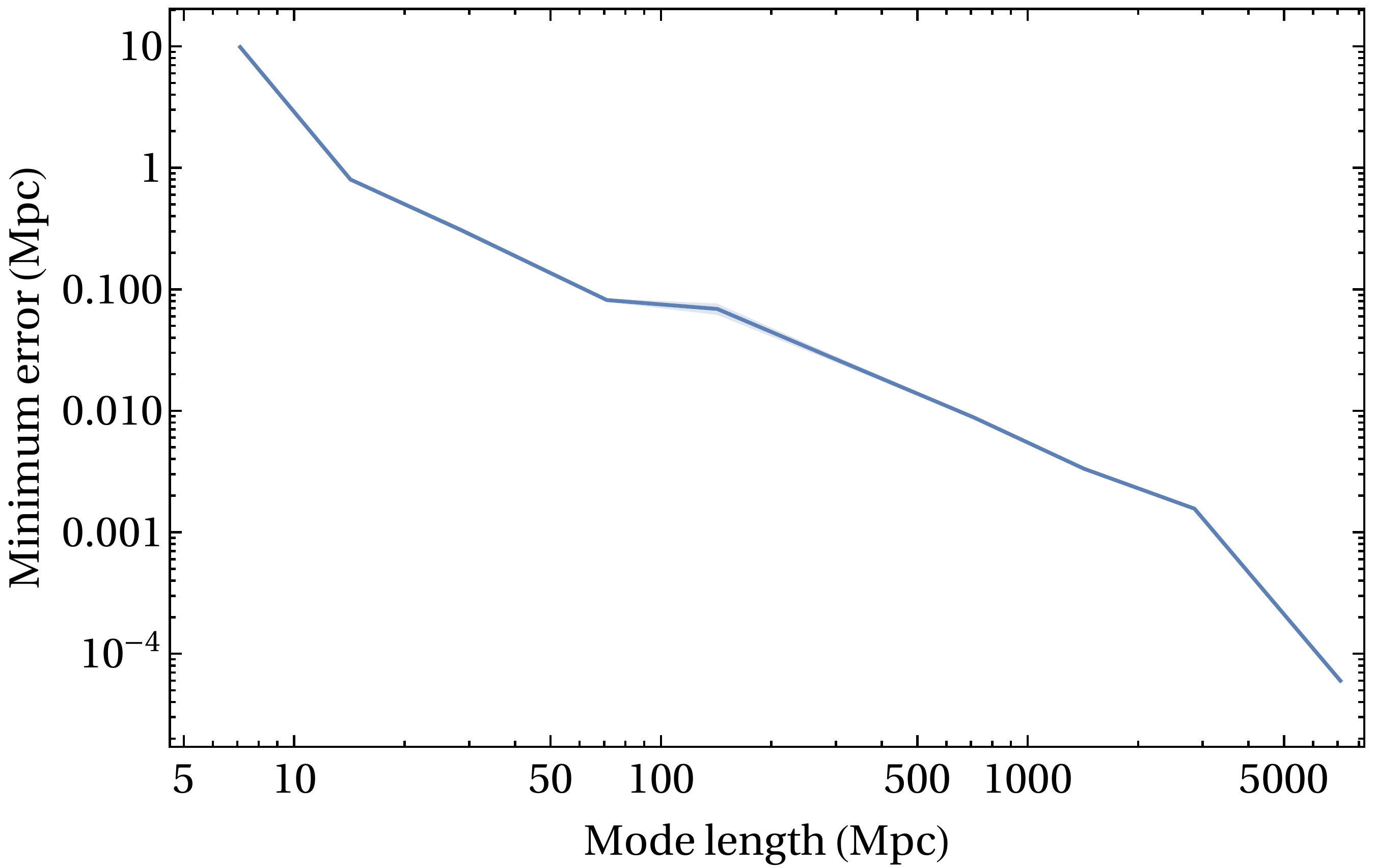}}
\caption[]{\label{fig:err_vs_k} Minimum constant errors in comoving \rm{Mpc} needed to validate the FLRW fittings (model M2) for different 
mode lengths.}
\end{figure}

Since the energy budget contains only matter and cosmological constant, we focus our particular attention on FLRW model M2, which has precisely those and only those two stress energy components.
Fig. \ref{fig:err_vs_k}
shows the minimum distance errors in the observational data points we would need to justify adding a third (spurious) stress-energy component to the FLRW fit to the Hubble diagram. 
We see that if we add a zero-mode (top row--infinite wavelength) fluctuation, it takes ridiculously small errors on observations to be able to notice that the two-component model is failing. 
As we decrease the wavelength of the perturbation, the less precision it takes to detect a mode of a given amplitude.

One should note that cosmological surveys measure the 
luminosity of supernovae to extract distance information, and the 
distance measure is quantified by the distance modulus 
$\mu = 5 \log_{10}{(D_L / 1 \rm Mpc)} + 25$. An error of $10 \, \rm Mpc$ (largest mock error needed for fitting to  $7\,\rm Mpc$ modes with model M2) corresponds to an error in $\mu$ of from $.003$ to $.05$ approximately, for supernovae ranging from $z = 1$ to $z=0.1$, 
with a mean error $\bar{\delta \mu} \approx 0.021$. These errors are an order of magnitude smaller than the estimated standard deviation of the error distribution $\sigma_{\mu} = 0.16$ given by LSST collaboration \cite{collaboration2009lsst}.
Therefore the  inhomogeneous modes we study will not bias 
the distance-redshift relation in a detectable way. 
In addition, the dominant contribution to the errors in the model M1 (the first column in Table \ref{tab:error_table}) is the homogeneous background, i.e., the inhomogenities from 
smaller mode other than 0-mode will not introduce extra errors for 
a fitting with a pure matter model. 
Therefore, the inhomogeneities will not bias our interpretation of  the cosmological constant as well.

Small inhomogeneities add curvature to the simulation; however, we show here that including $\Omega_k$ (model M3) in the fitting
{\sl will not reduce the error needed to validate the fitting}, except for, to a small degree, the two horizon-sized modes ($\lambda = H_I^{-1}$ and $\lambda = 0.5 H_I^{-1}$).
For these modes,
a smaller hypothetical error is needed for the model M3 compared to the model M2, 
indicating that a marginally better fit can be achieved.
For the mode $\lambda = 0.5 H_I^{-1}$, adding a hypothetical radiation component $\Omega_r$
would further improve the fitting. 
We argue that these additional effective matter components introduced by Hubble size modes is simply from fitting an oscillatory distance-redshift relation (see Fig. \ref{fig:avg}) with a polynomial-like model (see Eq. \eqref{eq:dl}). The fit can only be improved when the number of periods of the oscillation is comparable to the parameter number; a frequency that is too high or too low will not validate adding extra fitting parameters, and thus, considering small scale inhomogeneities will not introduce bias to extra matter components, including $\Omega_k$ or $\Omega_r$.

\subsection{Interpreting $\Omega_k$}
\label{sec:sub1}

We next investigate 
the connection between these $\Omega_k$ and averaged properties on spatial hypersurfaces or light-cones. 
We compare the best-fit $\Omega_k$ to $\Omega_R=-\left<R\right>/(6H_0^2)$ for all our observers, choosing two different averaging prescriptions for $\left<R\right>$. First, we compare to a volume-element-weighted average $\left<R\right>_{\rm Hyp.}$ on the spatial slice at
$z = 0$, which we notate as $\Sigma_0$. This average is performed over a circular coordinate patch centered on the observers, with a diameter given by the Hubble length scale at the time of the observation. We then compare to an average over the entire light cone surface, $\left<R\right>_{\rm LC}$, for each observer according to
\begin{align}
\label{eq:LC}
    \left<R\right>_{\rm Hyp.} & = \frac{\int_{\Sigma_0} d^3x \sqrt{\gamma} R  }{\int_{\Sigma_0} d^3x \sqrt{\gamma} } &
    \left<R\right>_{\rm LC} & = \frac{\int_{\rm LC} s^2 ds d\Omega  \sqrt{\gamma} (a^2 R)  }{\int_{\rm LC} s^2     ds d\Omega \sqrt{\gamma}},
\end{align}
where $s$ is the affine parameter (equivalent to the proper time in our case), and the light-cone integration is performed out to a time corresponding to a redshift of approximately $z=1$. Note that we manually add an extra term $a^2$ as the weight to the Ricci scalar $R$ in the light-cone integral in Eq. \eqref{eq:LC} to make sure that the equivelence between $\Omega_R$ and $\Omega_k$ is fully recovered under the FLRW limit.

\begin{figure}[htbp]
\centerline{\includegraphics[width=0.7\textwidth]{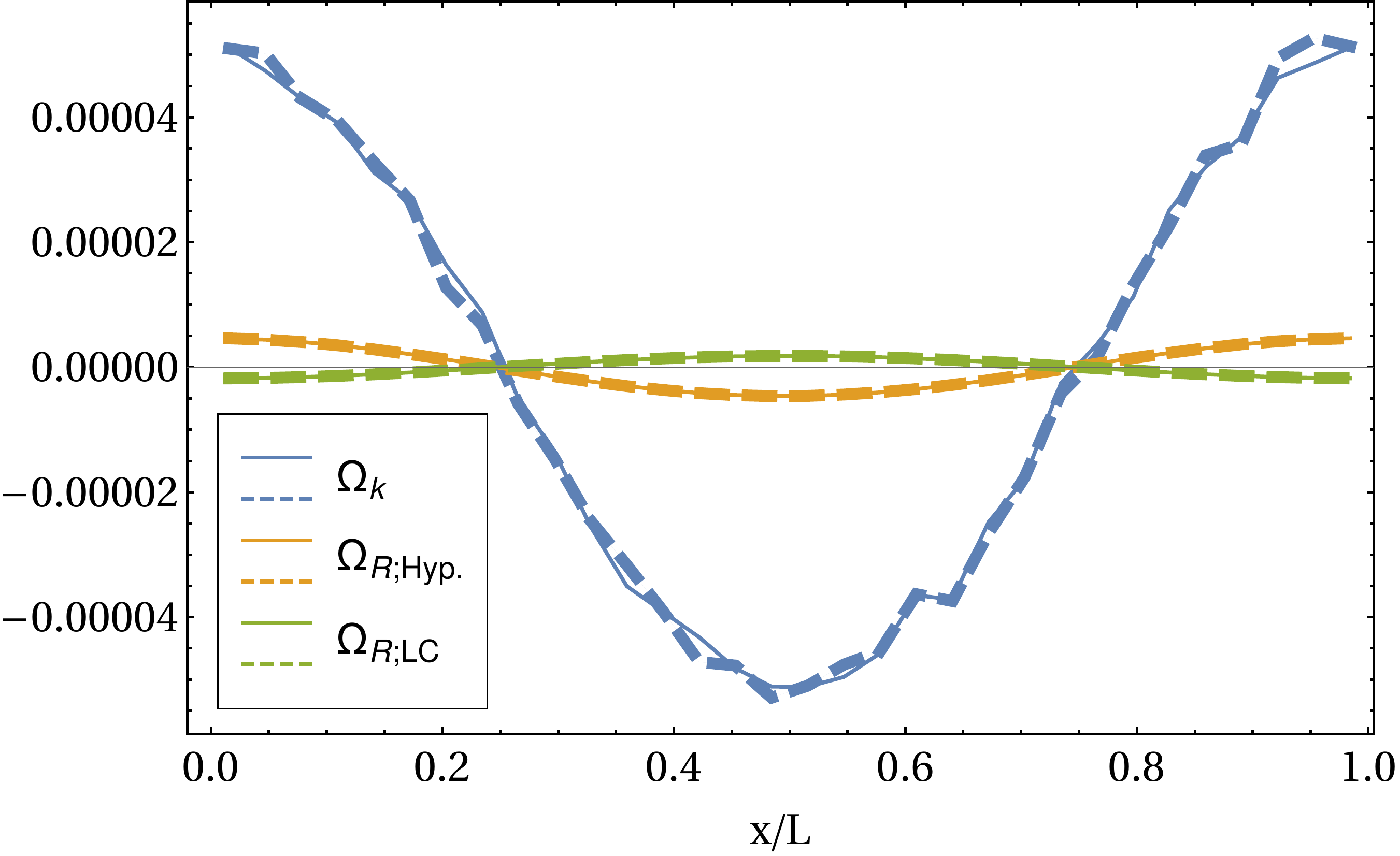}}
\caption[]{\label{fig:compare_omega} Comparison between $\Omega_k$ (blue) from the fitting to the FLRW model, 
$\Omega_R$ on the spatial hypersurface (orange), and $\Omega_R$ on the light cone (green) for observers 
standing along the x-axis.
Solid lines are exact results, while dashed lines roughly correspond to expectations from linear theory.
}
\end{figure}

Neither of these averages coincide with the best-fit
$\Omega_k$, as shown in Fig. \ref{fig:compare_omega} for a horizon size mode $\lambda = H_I^{-1}$. The numerical and sampling errors are not large here, indicating a clear difference: fluctuations in the fitted $\Omega_k$ values arise from the sampling variance (wiggles in the blue line), and
the numerical confidence interval of the green and orange lines are too small to be visible. The $\Omega_k$ found by averaging curvature over the hypersurface is found to be considerably smaller than the fit value; the lightcone-averaged curvature is even smaller, due to more heavily-weighted information at earlier times when density and curvature perturbations were not as large. Due to the symmetry of our setup, we do not ascribe any particular significance to the (anti-)correlations between the different $\Omega$s; rather the important conclusion we draw here is that they are not equivalent.

Nevertheless, these averages are still well-described by linear perturbation theory for the large-scale, small-amplitude mode shown.
We quantify the impact of non-linear effects
in Fig. \ref{fig:compare_omega}
by comparing the between exact, fully nonlinear results from
a mode with a baseline amplitude $A_0$ (solid lines), and a mode evolved with smaller (linearized) amplitude $A_0/10$ but scaled back up by 
multiplying a factor $10$ (dashed lines)\footnote{For any function $f(x,A)$, without any knowledge of its specific form, 
a Taylor expansion indicates that that the difference between $f(x, A+\delta A)$ and $ f(x, A+ c \delta A) / c$ vanishes exactly if the $f(x,A)$ is linear for a small $A$ and a constant $c$, or in the limit that $c \rightarrow 0$.
Any discrepancy between these two expressions quantifies the impact of non-linearities. Here we have chosen $c=10$}.
Any difference will indicate a contribution from non-linearities.
The overlap between solid and dashed lines therefore indicates the results are almost entirely explained by a linear, perturbative treatment.

\begin{figure}[htbp]
\includegraphics[width=0.49\textwidth]{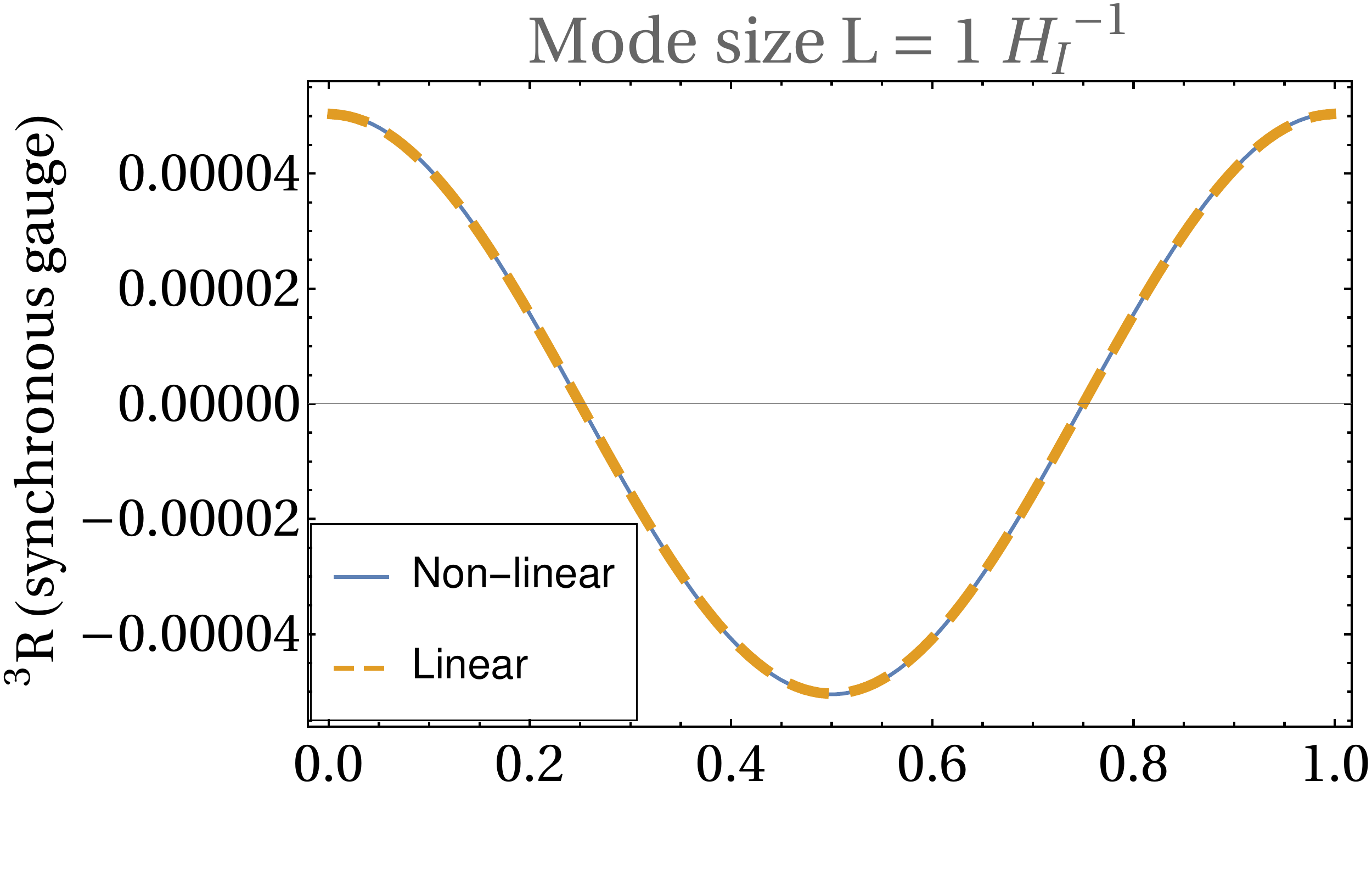}
\includegraphics[width=0.455\textwidth]{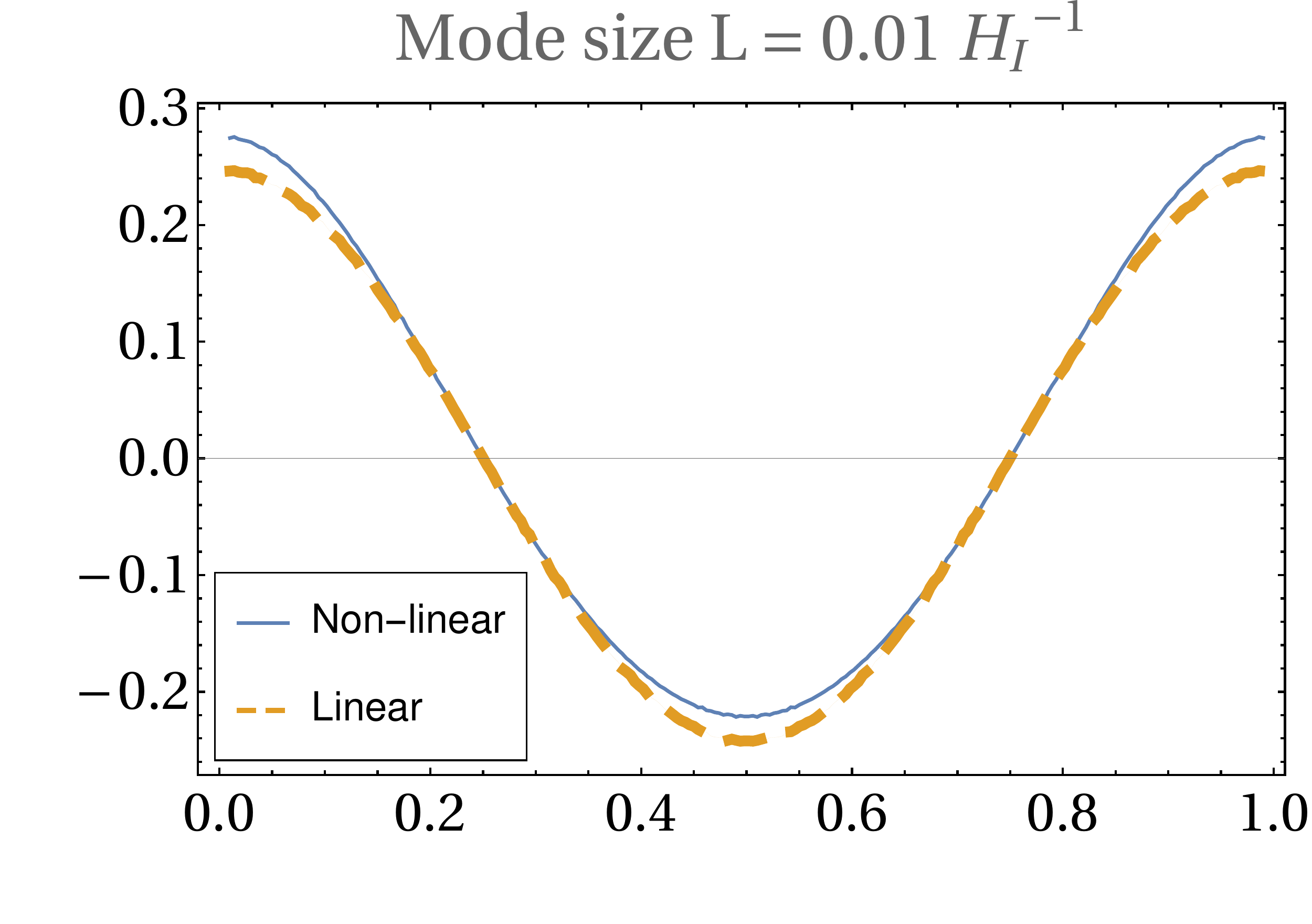}
\includegraphics[width=0.49\textwidth]{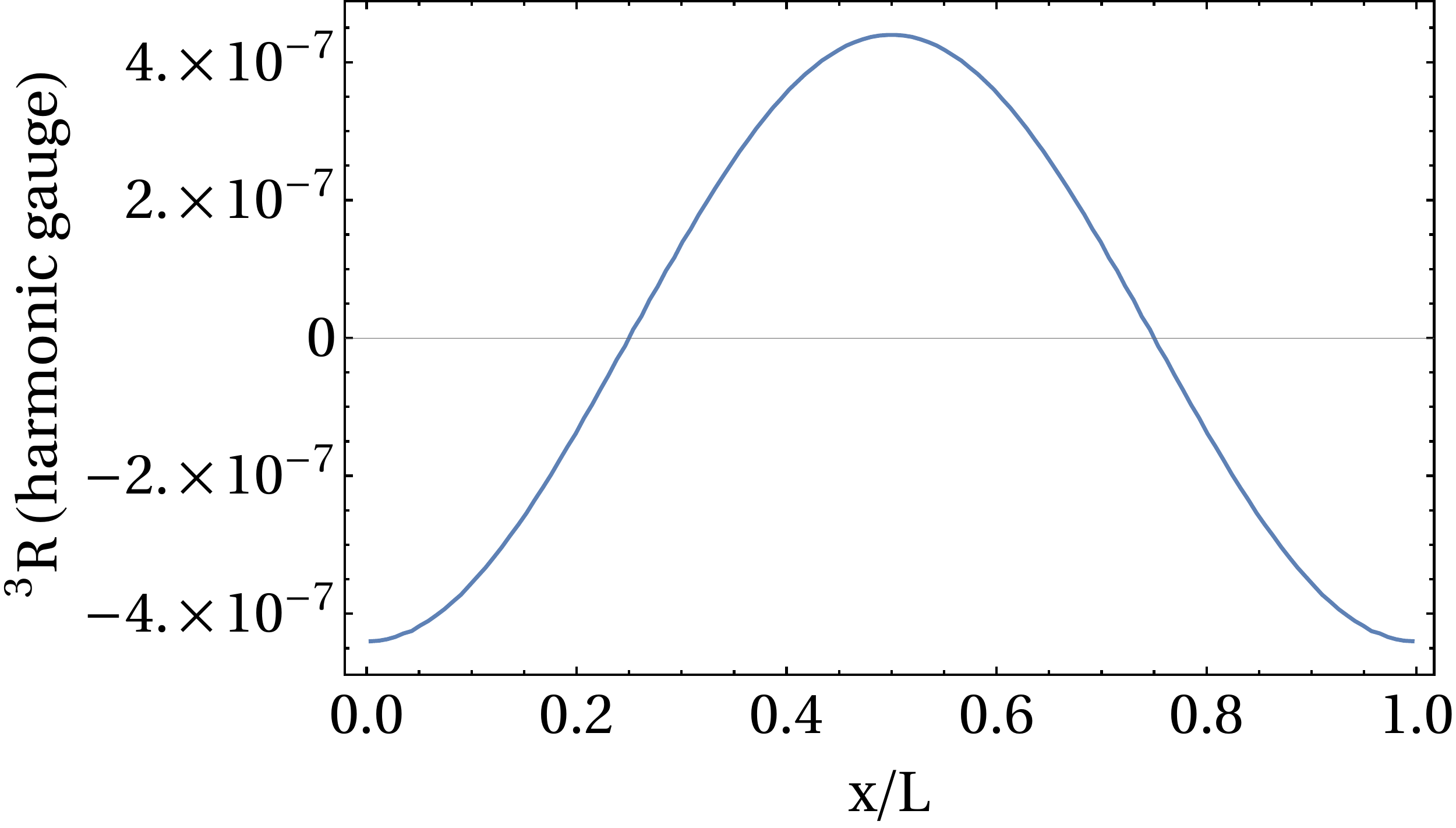}
\includegraphics[width=0.455\textwidth]{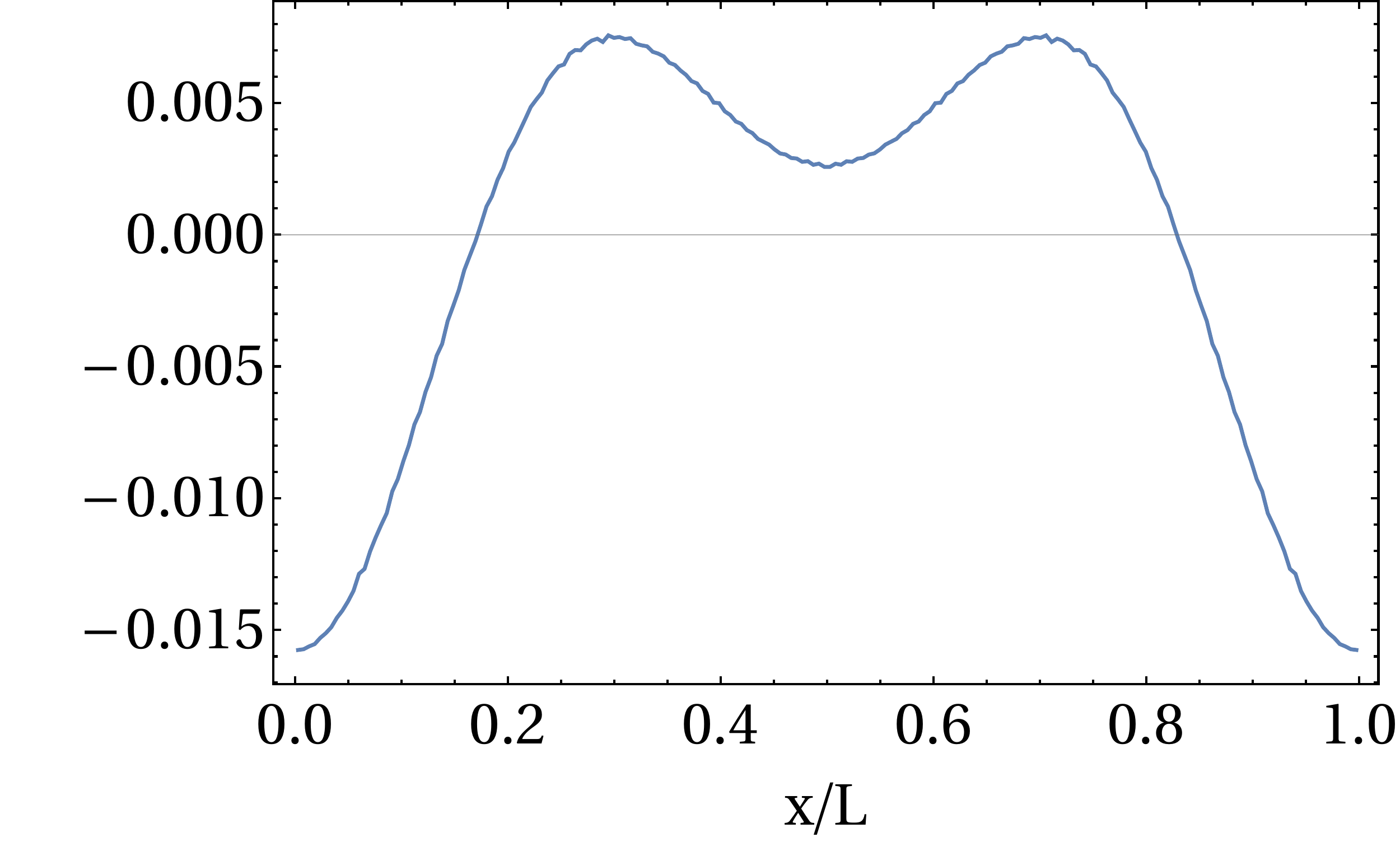}
\caption[]{\label{fig:non-linear} Comparison of the value of Ricci scalar $R$ on the $z=0$ hypersurface between $\lambda = H_I^{-1}$ mode (left)
and $\lambda = 0.01 H_I^{-1}$ mode (right). Solid lines are exact results, while dashed lines roughly correspond to expectations from linear theory.
The bottom sub-plots are presenting slices of $R$ calculated at the same redshift but 
with harmonic gauge.}
\end{figure}

It is then interesting to ask if this behavior persists on smaller scales, where mode amplitudes can be considerably larger, and nonlinearities are manifest; as well as to examine how the magnitude of curvature perturbations compare for different gauge choices.
Fig. \ref{fig:non-linear} shows a comparison of the curvature perturbation for $\lambda = H^{-1}$ and 
$\lambda = 0.01 H_I^{-1}$.
Non-linearities become apparent for the mode $\lambda = 0.01H_I^{-1}$ (and smaller): in this case, the curvature is not well-described by linear perturbation theory.
The bottom panel in Fig. \ref{fig:non-linear} then shows the Ricci scalar at the same redshift, but in
harmonic slicing\footnote{The harmonic slicing condition we use is 
\begin{align}
    \partial_t \alpha = \alpha^2 (K-\left<K\right>), \;\;\;\; \beta =0,
\end{align}
where $\left<K\right>$ is the averaged extrinsic curvature on spatial hypersurfaces.}. 
The order of magnitude differences indicate the averaged Ricci scalar will strongly depend on the gauge choices, and in either case will not coincide with the best-fit $\Omega_k$.

\section{Conclusion}

We have quantified the possible deviations from the FLRW model introduced 
by inhomogeneous perturbations through analyzing the simulated Hubble diagram 
measured by comoving observers. We have shown that realistic amplitudes
of perturbations will not bias the distance-redshift relation to a detectable extent. 
The inhomogeneities only introduce an effective $\Omega_k$ for
a Hubble size mode, and the value of $\Omega_k$ is small and proportional 
to the amplitude of the perturbation. 
Therefore, our numerical results strongly suggest that the 
cosmic inhomogeneities will not introduce detectable deviations, and 
curvature perturbations across a broad range of dynamical scales cannot bias a measurement of the curvature component $\Omega_k$, unless with an amplitude inconsistent with standard inflationary initial conditions within $\Lambda$CDM.

We have also investigated carefully the Hubble size modes that are able 
to induce an effective $\Omega_k$. It has shown that the value of $\Omega_k$ is different than 
what would be expected from the average of the three-dimensional Ricci-curvature on the spatial hypersurface or the light-cone, whose evolution can become non-linear when going to the smaller scales. This study also provides further evidence for the discrepancy between spatial and volume averages in curved spacetimes, which has been discussed
in e.g. \cite{Lavinto:2013exa, Koksbang:2019wfg} in the context of the Swiss cheese 
model.

We can qualitatively compare to other work that examines deviations from the FLRW model or biases in the inferred curvature due to inhomogeneity in a relativistic setting. Meures {\it et al}. \cite{Meures:2011gp} analyzes the averaged behavior of redshifts and distances observables along lines of sight with an exact inhomogeneous GR solution. However, only a sub-percent level of deviation from the FLRW model is identified when the fluctuation wavelength is as large as $500\,\rm Mpc$, which is consistent with our work.
Adamek {\it et al}. \cite{Adamek:2018rru} performs a large scale N-body simulation, and includes a more realistic distribution of matter across a wider range of scales.
Next to this study, the effects we find turn out to be too small to bias the FLRW fitting, and a non-zero $\Omega_k$ will not be interpreted from the fittings except for cases containing Hubble-size perturbations.
Our result is instead complementary to \cite{Adamek:2018rru}, covering shorter, moderately nonlinear modes up to Hubble-scale modes; but importantly, the fully general relativistic treatment allows us to examine the discrepancy between $\Omega_k$ and $\Omega_R$.

To summarize, our numerical experiment reinforces the robustness 
of the FLRW mode when fitting cosmological survey data,
which is in agreement with other numerical investigations 
with full general relativity \cite{Giblin:2017ezj,Bentivegna:2015flc,Giblin:2015vwq,Macpherson:2018akp} but with a focus on cosmological 
observables.
The difference between 
cosmic observables $\Omega_k$ and averaged curvature information on hypersurface or light-cone is confirmed.

\section{Acknowledgements}

This work made use of the High Performance
Computing Resource in the Core Facility for Advanced Research Computing
at Case Western Reserve University.
JTG is supported by the National Science Foundation Grant No. PHY-2013718.
GDS and CT were supported in part by grant DE-SC0009946 from the US DOE. 
SA was supported in part by the project 
``Combining Cosmic Microwave Background and Large Scale Structure data: an Integrated Approach for Addressing Fundamental Questions in Cosmology'',  funded  by  the  MIUR  Progetti  di  Rilevante  Interesse  Nazionale (PRIN) Bando 2017 - grant 2017YJYZAH.

\bibliography{references}

\appendix

\section{Code validation and Convergence test}

\begin{figure}[htbp]
\centerline{\includegraphics[width=0.75\textwidth]{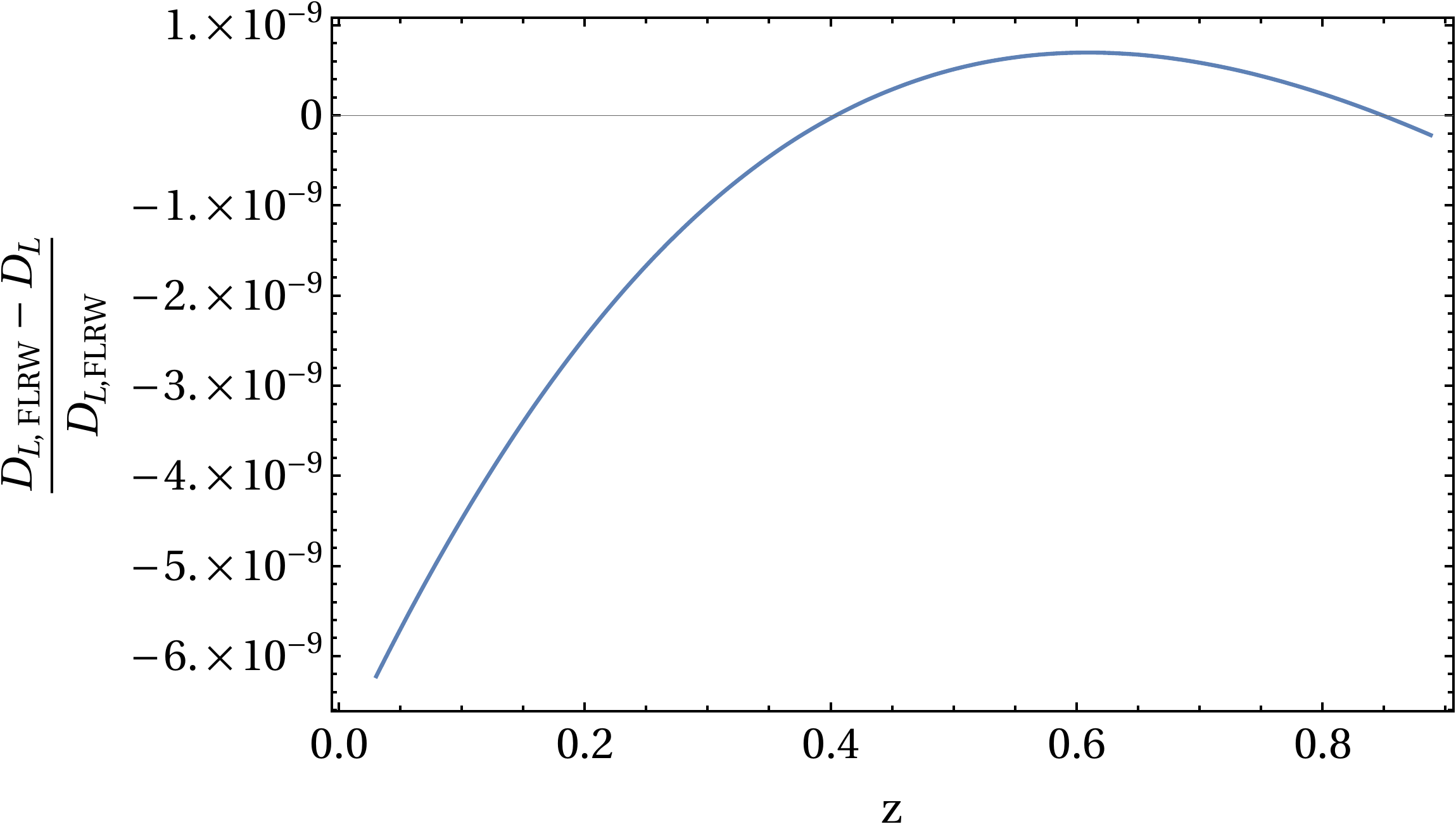}}
\caption[]{\label{fig:FLRW_test_residual} The relative fitting residual to the best fitted FLRW model for a zero-mode test. The box size is $L = 0.01 H^{-1}$}
\end{figure}

Here we present several code validation tests. We first examine a pure zero-mode solution, or an exactly homogeneous 
universe. In Fig. \ref{fig:FLRW_test_residual} we find only a very small numerical deviation from the FLRW model, indicating a high degree of numerical precision.

\begin{figure}[htbp]
\centerline{\includegraphics[width=0.7\textwidth]{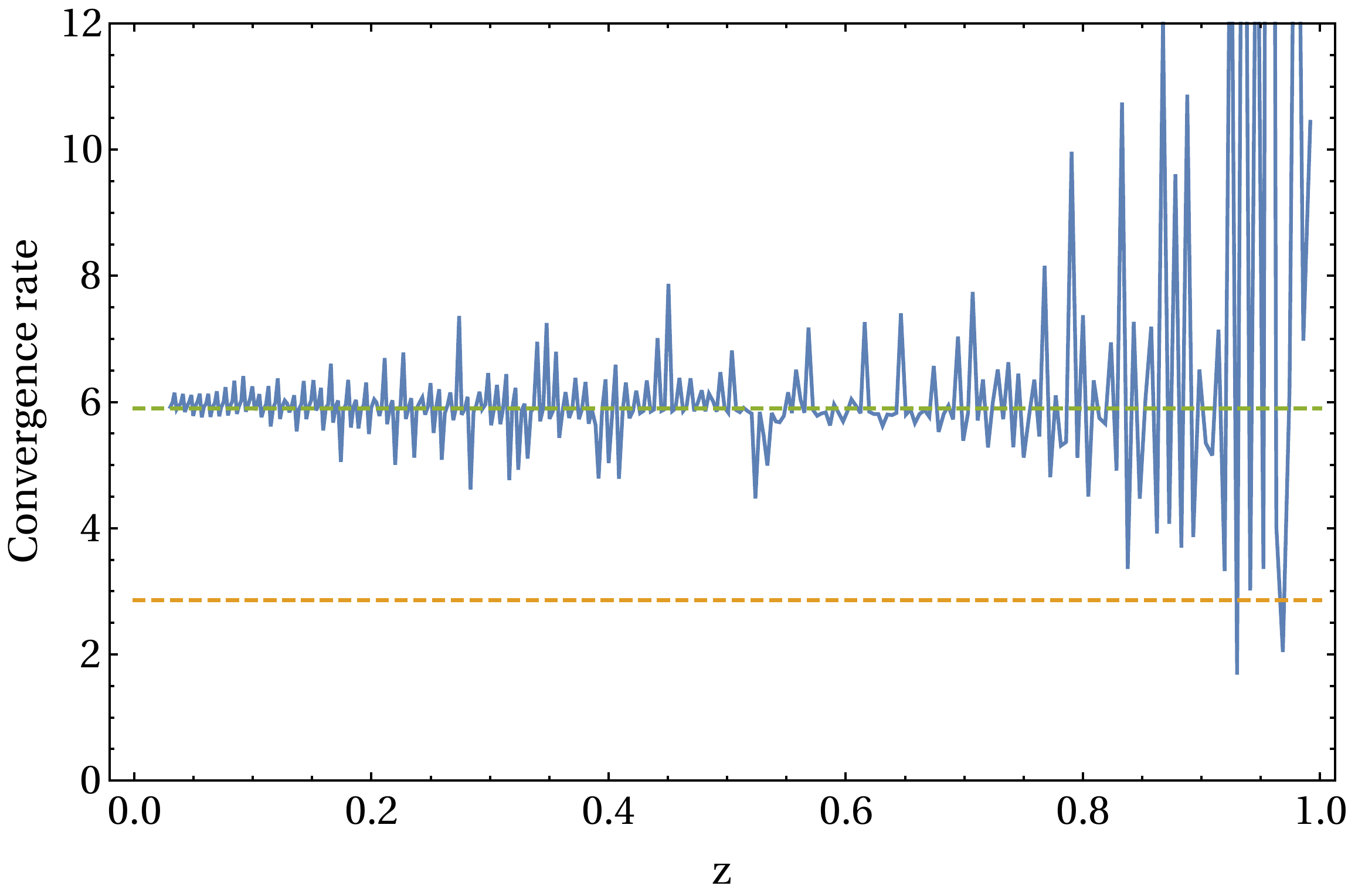}}
\caption[]{\label{fig:conv} The convergence rate for the light ray that has the worst convergence rate: a polar angle around 45 degrees for the observer standing at the largest density contrast point. 
This is also for the smallest mode we investigate, $\lambda = 0.005H_I^{-1}$. The blue and orange dashed line 
indicate 2nd and 4th order convergence respectively. Approximately 4th order convergence is 
achieved as expected.}
\end{figure}

We also validate results for all of our simulations by checking the numerical convergence rate for the distance-redshift relation for each run. 
The convergence rate is calculated as
\begin{eqnarray}
\label{eq:c}
 c \equiv \frac{|f_{N_c} - f_{N_m}|}{|f_{N_m} - f_{N_f}|},
\end{eqnarray}
where $f_{N_c}$, $f_{N_m}$ and $f_{N_f}$ are values calculated at resolutions $N_c$, $N_m$ and $N_f$, which are coarsest to finest.  Resolutions are chosen to be $128$, $192$ and $256$ respectively in our tests. 
The slowest convergence happens with the smallest mode ($\lambda = .005H^{-1}$) and 
rays with polar angle near $45$ degrees. In Fig. \ref{fig:conv} we show the convergence rate for such 
a ray, which indicates 4th order convergence is achieved as expected. 

\section{Extracting distances information in GR}

To extract distance information during ray-tracing, we need to track the Jacobian
matrix $D^A_B$, which encodes information about image distortion. The Jacobian evolves according to
\begin{align}
\label{eq:6}
\nabla_{p} \nabla_{p} D_{\;\;B}^A = R^A_{\;\;C \mu\nu} p^{\mu} p^{\nu} D^{C}_{\;\;B},
\end{align}
which is derived from the geodesic deviation equation (GDE).
However, this equation cannot be incorporated into the 3+1 scheme employed
by numerical relativity easily due to the difficulty of extracting Ricci tensor
on the full 4-dimensional manifold. We instead use a technique in which we
trace the infinitesimal area on the observer's screen.

First, after initializing a main ray with 4-vector $p_0^{\mu}$, two adjacent auxiliary rays with 4-momentum 
$p_1= p_0^{\mu} + \epsilon \hat{s}_1^{\mu}$ and $p_2^{\mu} = p_0^{\mu} + \epsilon \hat{s}_2^{\mu}$ are also initialized. Infinitesimal separation vectors can be defined as
\begin{align}
\label{eq:9}
  \xi_1^{\mu} &= p_1^{\mu} - p_0^{\mu} \\
  \xi_2^{\mu} &= p_2^{\mu} - p_0^{\nu},
\end{align}
and their evolution is governed by the GDE.

We can then define an observer's ``screen space'', which
is a 2-dimensional plane. 
The operator that projects 4-vectors to spatial vectors to
the observer's ``screen space'' can be written as
\begin{align}
\label{eq:7}
S_{\mu\nu} = g_{\mu\nu} + u_{\mu} u_{\nu} - d_{\mu} d_{\nu},
\end{align}
where $u^{\mu}$ is the observer's 4-velocity and $d_{\mu}$ is the
observer's light of sight. The light of sight vectors can be defined
by projecting the 4-velocity of photons into observers's frame through
\begin{align}
\label{eq:10}
  d_{\nu} \equiv \frac{1}{\omega} P_{\;\;\nu}^{\mu} p_{\mu}
  = \frac{1}{\omega} p_{\nu} - u_{\nu}.
\end{align}
The last equality provides the definition of observer's projector:
\begin{align}
\label{eq:11}
P_{\mu\nu} = g_{\mu\nu} - u_{\mu} u_{\nu}.
\end{align}

Two screen orthogonal basis can be chosen by projecting two arbitrary non-parallel vectors
$V_1^{\mu}$ and $V_2^{\mu}$,

\begin{align}
\label{eq:8}
  s_{\mu}^{1}&=\frac{S_{\mu\nu}V_{1}^{\nu}}{\sqrt{S_{\mu\nu}V_{1}^{\mu}V_{1}^{\nu}}} \\
  s_{\mu}^{2}&=\frac{S_{\mu\nu}V_{2}^{\nu}-s_{\mu}^{1}s_{\nu}^{1}V_{2}^{\nu}}{\sqrt{S_{\mu\nu}V_{2}^{\mu}V_{2}^{\nu}-s_{\mu}^{1}V_{2}^{\mu}s_{\nu}^{1}V_{2}^{\nu}}}\,.
\end{align}
In practice, we get the the screen basis vectors by projecting the separation vectors
$\xi_1$ and $\xi_2$.

After acquiring the screen basis and separation vectors, an infinitesimal area element
can be written as
\begin{align}
\label{eq:12}
A={\rm det}\left(\begin{array}{cc}
s_{\mu}^{1}\xi_{1}^{\mu} & s_{\mu}^{2}\xi_{1}^{\mu}\\
s_{\mu}^{1}\xi_{2}^{\mu} & s_{\mu}^{2}\xi_{2}^{\mu}
\end{array}\right)
\end{align}

And, finally, the luminosity distance is
\begin{align}
\label{eq:13}
D_L = (1+z)^2\sqrt{\frac{A}{\Omega}}, 
\end{align}
where the solid angle of the observer $\Omega$ simply equals to $\epsilon^2$ to the first order in $\epsilon$.

\label{appd:2}

\end{document}